\documentclass[12pt, prd, showpacs]{revtex4}
\usepackage{graphicx}
\usepackage{amssymb}
\usepackage{amsmath}

\begin{document}

\title{Regular frames for spherically symmetric black holes revisited}
\author{A. \ V. Toporensky}
\affiliation{Sternberg Astronomical Institute, Lomonosov Moscow State University}
\affiliation{Kazan Federal University, Kremlevskaya 18, Kazan 420008, Russia}
\email{atopor@rambler.ru}
\author{O. B. Zaslavskii}
\affiliation{Department of physics and Technology, Kharkov V.N. Karazin National
University, 4 Svoboda Square, Kharkov 61022, Ukraine\\
Kazan Federal University, Kremlevskaya 18, Kazan 420008, Russia}
\email{zaslav@ukr.net}

\begin{abstract}
We consider a space-time of a spherically symmetric black hole with one
simple horizon. As a standard coordinate frame fails in its vicinity, this
requires continuation across the horizon and constructing frames which are
regular there. Up to now, several standard frames of such a kind are known.
It was shown in literature before, how some of them can be united in one
picture as different limits of a general scheme. However, some types of
frames (the Kruskal-Szekeres and Lema\^{\i}tre ones) and transformations to
them from the original one remained completely disjoint. We show that the
Kruskal-Szekeres and Lema\^{\i}tre frames stem from the same root. Overall,
our approach in some sense completes the procedure and gives the most
general scheme. We relate the parameter of transformation $e_{0}$ to the
specific energy of fiducial observers and show that in the limit $%
e_{0}\rightarrow 0$ a homogeneous metric under the horizon can be obtained
by a smooth limiting transition.
\end{abstract}

\keywords{frame, black \ hole, coordinate transformations}
\pacs{04.70.Bw, 97.60.Lf }
\maketitle

\section{Introduction}

The Schwarzschild black hole \cite{schw} is the core object of general
relativity, the properties of its space-time play a crucial role in
understanding space-time of black holes. The standard coordinate system in
which the Schwarzschild metric is written uses so-called curvature
(Schwarzschildean) coordinates and fails on the event horizon. \ To repair
this drawback, there are several "standard" transformations and
corresponding coordinate systems - such as Eddington-Filkestein coordinates
(EF), Kruskal-Szekeres ones, Novikov systems, Gullstrand-Painlev\'{e} (GP)
and Lema\^{\i}tre coordinates. All these coordinates and methods of their
constructions look very different. Meanwhile, it turned out that all of
these transformations (or at least their part) can be united, if one
introduces an additional parameter in the coordinate transformation. This
parameter $e_{0}$ has the meaning of the energy per unit mass for a
reference (fiducial) observer. In this sense, particle dynamics is encoded
in the typical transformations to the regular frame. Then, taking the limit $%
e_{0}\rightarrow \infty $, one can recover some familiar coordinate systems
and metrics \cite{mart}, \cite{finch}, \cite{jose}. In this sense, previous
metrics are contained as different limiting cases of some more general one.

The approach developed in \cite{mart}, \cite{jose} does not include the
transformation to the synchronous frame. Meanwhile, this frame simplifies
the whole picture and thus plays an important role. The construction of such
a frame for the Schwarzschild metric was done by Lema\^{\i}tre \cite{lem}.
Quite recently, this was generalized to metrics other than the Schwarzschild
one. To this end, two different procedures were suggested, \cite{bron}, \cite%
{3}.

In spite of the fact that some frames were combined in a single general
construction, the whole picture remains quite intricate and even the ways of
particular unifications also look very different. Unifying particular
approaches and metrics, we can separate the whole set of possible
transformations to two kinds. The first one (A) envolves the parameter $%
e_{0} $ having the meaning of energy per unit mass of fiducial observers.
This includes such systems in which fiducial observers (characterized by a
constant value of a spatial coordinate) move not freely. The bright example
is the Kruskal-Szekeres system. The second class (B) contains the transition
to synchronous systems. For example, this concerns the Novikov system \cite%
{nov63} or more general Bronnikov - Dymnikova - Galactionov (BDG) system 
\cite{bron}. In appearance, classes A and B look completely separated,
derived from different requirements and seem to be complementary to each
other. However, we show that, as a matter of fact, there is deep and very
simple connection between both classes. We also consider the limit quite
different from \cite{mart}, \cite{finch}, \cite{jose} where it was implied
that $e_{0}>1$. This is the limit $e_{0}\rightarrow 0$. Then, another
synchronous system typical of the Kantwoski-Sachs (KS) metric \cite{komp}, 
\cite{ks} appears explicitly.

The most general regular frame is the one suggested by Fomin \cite{fom}.
Wrongly, this paper was not noticed in due course and remained poorly cited.
We use the approach of Fomin to show that all other ones can be obtained
from it.

Our consideration applies to a class of metrics more general than the
Schwarzschild one. It includes the Schwarzschild-de Sitter, Reissner-Nordstr%
\"{o}m metrics, etc. Further generalization is straightforward. We use the
geometric system of units in which fundamental constants $G=c=1$.

\section{Generalized Gullstrand-Painlev\'{e} frame}

Let us consider the metric%
\begin{equation}
ds^{2}=-fdt^{2}+\frac{dr^{2}}{f}+r^{2}d\omega ^{2}\text{,}  \label{met}
\end{equation}%
where $d\omega ^{2}=d\theta ^{2}+d\phi ^{2}\sin ^{2}\theta $. It represents
the spherically symmetric solution of the Einstein equations, provided the
components of the stress-energy tensor obey the relation $%
T_{r}^{r}=T_{0}^{0} $. Let $r=r_{+}$ correspond to the event horizon, so $%
f(r_{+})=0.$ The original frame fails in the vicinity of the horizon. To
repair the situation, one can introduce a new time variable

\begin{equation}
d\tilde{t}=e_{0}dt+e_{0}\frac{dr}{f}V\equiv e_{0}dt+P_{0}\frac{dr}{f}\text{,}
\label{tmp}
\end{equation}%
where initially we consider $e_{0}$ as a positive function of coordinates $r$
and $t$, 
\begin{equation}
P_{0}=e_{0}V=\sqrt{e_{0}^{2}-f},  \label{p0}
\end{equation}%
\begin{equation}
V\equiv \sqrt{1-\frac{f}{e_{0}^{2}}}\text{.}  \label{V}
\end{equation}

After substitution in (\ref{met}), one obtains%
\begin{equation}
ds^{2}=-\frac{f}{e_{0}^{2}}d\tilde{t}^{2}+\frac{2d\tilde{t}dr}{e_{0}}V+\frac{%
dr^{2}}{e_{0}^{2}}+r^{2}d\omega ^{2}\text{.}  \label{GP}
\end{equation}%
It can be also written in the form%
\begin{equation}
ds^{2}=-d\tilde{t}^{2}+\frac{1}{e_{0}^{2}}(dr+Ve_{0}d\tilde{t}%
)^{2}+r^{2}d\omega ^{2}\text{.}  \label{gp1}
\end{equation}

Now, 
\begin{equation}
g^{\tilde{t}\tilde{t}}=-1\text{, }g^{r\tilde{t}}=Ve_{0}\text{, }g^{rr}=f.
\label{kontr}
\end{equation}%
As $d\tilde{t}$ should be a total differential, the integrability conditions
have to be fulfilled:%
\begin{equation}
\left( e_{0}\right) _{,r}=\frac{e_{0}\left( e_{0}\right) _{,t}}{P_{0}f}\text{%
.}  \label{intrt}
\end{equation}

Eq. (\ref{GP}) corresponds to eq. (3.19) of \cite{finch}. Let $e_{0}=const$.
If $e_{0}=1$, we arrive at the Gullstrand-Painlev\'{e} frame \cite{gul}, 
\cite{p}, generalized to an arbitrary $f$ \cite{3}. If $f=1-\frac{2m}{r}$ in
(\ref{GP}), we return to the Schwarzschild version of the GP system \cite%
{lake}, \cite{mart}, \cite{jose}.

For the GP system the cross term with $d\tilde{t}dr$ defines a coordinate
flow velocity $V$ which has a direct physical meaning - it is the 3-velocity
of a free falling particle with the unit energy with respect to the static
coordinate system (see below). From (5) we can see that this direct
correspondence is lost since for $e_{0}\neq 1$, the additional factor $%
(1/e_{0})$ appears. The reason is, however, rather clear -- since the
intervals of physical distance $dl$ in $\tilde{t}=const$ sections are
connected with the interval of coordinate distance $dr$ via $dl=dr/e_{0}$,
the physical 3-velocity of the generalized GP system with respect to the
stationary system is still equal to $V$, as it should be. So, we still can
think of the coordinate system with $e_{0}=const$ as realized by free
falling particles with the energy $e_{0}$.

This can be confirmed by direct simple calculations. We can choose tetrads
attached to an static observer. Then, in coordinates $(t,r,\theta ,\phi )$%
\begin{equation}
h_{(0)\mu }=\sqrt{f}(-1,0,0,0),
\end{equation}%
\begin{equation}
h_{\left( 1\right) \mu }=(0,\frac{1}{\sqrt{f}},0,0).
\end{equation}%
Let us introduce the three-velocity in a standard way \cite{72}%
\begin{equation}
V^{(i)}=-\frac{h_{(i)\mu }u^{\mu }}{h_{(0)\mu }u^{\mu }}\text{,}  \label{vu}
\end{equation}%
where $u^{\mu }$ is the four-velocity. Eq. (\ref{vu}) is valid in general.
Now, we will apply it to motion of a particle moving freely with a constant
specific energy $e_{0}$. For pure radial motion,

\begin{equation}
\frac{dt}{d\tau }=\frac{e_{0}}{f}\text{,}  \label{t}
\end{equation}%
\begin{equation}
\frac{dr}{d\tau }=-P_{0}\text{,}
\end{equation}%
where $\tau $ is the proper time along the trajectory, $P_{0}$ is given by (%
\ref{p0}). Then, $u^{\mu }=(\frac{e_{0}}{f},-P_{0},0,0)$. By substitution
into (\ref{vu}), we obtain that%
\begin{equation}
V^{(1)}=V\text{.}
\end{equation}

Thus $V$ has the meaning of a velocity measured by a static observer, $%
P_{0}=e_{0}V$ being the corresponding momentum. Then, $e_{0}$ can be thought
of as the energy of some effective particle moving in the given background,%
\begin{equation}
e_{0}=\frac{\sqrt{f}}{\sqrt{1-V^{2}}}\text{, }P_{0}=\frac{V\sqrt{f}}{\sqrt{%
1-V^{2}}}\text{.}  \label{e0}
\end{equation}

If a particle moves not freely, $e_{0}$ ceases to be an integral of motion
and depends on time. However, as stressed in Sec. 3 of \cite{our}, even in
this case equations of motion retain their validity with $e_{0}=e_{0}(t)$.
Moreover, we can admit formally dependence of spatial coordinates as well in 
$e_{0}$ that enters transformation (\ref{tmp}). Although $e_{0}$ is not an
integral of motion in this case, formulas (\ref{e0}) show that this can be
still considered as a pure local Lorentz transformation. Below we will see
how this helps in constructing regular frames.

The case of $e_{0}=1$ is a special one since spatial sections $d\tilde{t}=0$
are flat. In this case spatial intervals are simply differences $r_{2}-r_{1}$%
. When $e_{0}\neq 1$, the factor $1/e_{0}^{2}$ before $dr^{2}$ makes them
non-flat. For $e_{0}\geq 1$, the proper distance between points with fixed $%
r_{2}$ and $r_{1}$ measured along the hypersurface $\tilde{t}=const$ is less
than in the case when $e_{0}=1$. This is some reminiscent of the Lorentz
contraction in special relativity (SR). Indeed, in SR $(f=1$) the proper
distance along the hypersurface $t=const$ $(e_{0}=1$, $V=0)$ is equal to $%
\Delta r=r_{2}-r_{1}$. If another observer passes by him with velocity $V$
and the specific energy $e_{0}=\frac{1}{\sqrt{1-V^{2}}}>1$, the proper
distance measured in its own frame $(\tilde{t}=const)$ equals $\Delta l=%
\frac{\Delta r}{e_{0}}=\Delta r\sqrt{1-V^{2}}<\Delta r$. \ However, if $%
e_{0}<1$, the corresponding situation has no analogue in SR since along the
surface $\tilde{t}=const$ the proper distance $\Delta l>\Delta r$. This is
due to the fact that the space-time is curved since in the flat one such
observers are absent. Meanwhile, for any fixed $r$ there is a lower bound on
possible $e_{0}$ which is equal to $f(r)$. Therefore, among all states with
different $e_{0}$ and fixed $dr,$ the maximum value of the proper distance $%
dl=dr/e_{0}$ is achieved for a minimum value of $e_{0}=\sqrt{f(r)}$ that
coincides with the distance in the static frame. This is natural since an
observer with minimal possible $e_{0}$ for a given $r$ has zero flow
velocity (\ref{V}) and thus coincides with a stationary observer. In this
sense, there is some analogy with SR again since the minimum of the proper
distance is achieved in the rest frame.

If we rescale time according to 
\begin{equation}
\tilde{t}=\hat{t}e_{0}  \label{te}
\end{equation}%
with $e_{0}=const>0$,%
\begin{equation}
ds^{2}=-fd\hat{t}^{2}+2d\hat{t}drV+\frac{dr^{2}}{e_{0}^{2}}+r^{2}d\omega
^{2}.  \label{gp}
\end{equation}

It follows from (\ref{tmp}), (\ref{te}) that%
\begin{equation}
\hat{t}=t-\int \frac{dr}{f}V=t-\int dr^{\ast }V\text{,}
\end{equation}%
where 
\begin{equation}
r^{\ast }=\int \frac{dr}{f}  \label{tort}
\end{equation}%
is the tortoise coordinate.

If $f=1-\frac{2M}{r}$ and we write $e_{0}=\frac{1}{p^{2}}$, we return to the
coordinates of Ref. \cite{mart} - see eq. (3.5) there. It is worth noting
that eq. (\ref{GP}) is valid even if $e_{0}$ depends on both coordinates.
Meanwhile, transformation from (\ref{GP}) to (\ref{gp}) implies that $%
e_{0}=const$.

If one takes the limit $e_{0}\rightarrow \infty $ in (\ref{gp}), the metric
in the EF coordinates is reproduced:%
\begin{equation}
ds^{2}=-fd\hat{t}^{2}+2d\hat{t}dr+r^{2}d\omega ^{2}\text{.}
\end{equation}

In doing so, $V\rightarrow 1$, 
\begin{equation}
\hat{t}=t-r^{\ast }=u
\end{equation}%
is the EF coordinate.

It is worth noting that the transformation (\ref{tmp}) somewhat generalizes
that in \cite{jose}. However, in contrast to \cite{jose}, we do not use the
double GP coordinates and obtain the limit $e_{0}\rightarrow \infty $
directly from (\ref{GP}) after rescaling the time coordinate $\tilde{t}%
\rightarrow \hat{t}$. If one starts with $\hat{t}$ from the very beginning,
the limiting transition from \cite{mart} is reproduced easily with their $T$
equal to $\hat{t}$. Meanwhile, as the transformation used in \cite{mart}
does not include the parameter $e_{0}$, it is relatively restricted in that
it is unable to describe the diversity of different approaches.

\section{Diagonal metric}

We can consider transformation that can be interpreted as modification of
the approach developed by Fomin \cite{fom}. For completeness, we present
here the main features of this approach, though in contrast to the original
paper, we use our notations with the parameters $e_{0}$ and $P_{0}$.

Let in new coordinates $T$, $\rho $ the metric be regular and diagonal,%
\begin{equation}
ds^{2}=-F(T,\rho )dT^{2}+G(T,\rho )d\rho ^{2}+r^{2}(\rho ,T)d\omega ^{2}%
\text{.}  \label{fom}
\end{equation}

We perform transformations according to which%
\begin{equation}
dt=\frac{e_{0}}{f}\sqrt{F}dT-\frac{\sqrt{G}P_{0}}{f}d\rho ,  \label{dt}
\end{equation}%
\begin{equation}
dr=e_{0}\sqrt{G}d\rho -P_{0}\sqrt{F}dT\text{,}  \label{dr}
\end{equation}%
where $P_{0}$ is given by eq. (\ref{p0}).

The inverse transformation reads%
\begin{equation}
dT=\frac{1}{\sqrt{F}}(dte_{0}+\frac{dr}{f}P_{0})\text{,}  \label{T}
\end{equation}%
\begin{equation}
d\rho =\frac{1}{\sqrt{G}}\left( dtP_{0}+\frac{dre_{0}}{f}\right) .
\label{ro}
\end{equation}

It is easy to check that in new coordinates the metric is indeed diagonal.
If we put in (\ref{T}) $F=1$ and, instead of $\rho $, will use our previous
coordinate $r$, this would correspond to the transformation (\ref{tmp}) that
leads to the GP metric (\ref{GP}).

It follows from (\ref{dt}) - (\ref{ro}) that%
\begin{equation}
\frac{r_{,T}}{r_{,\rho }}\sqrt{\frac{G}{F}}=-V\text{,}  \label{rv}
\end{equation}%
\begin{equation}
F=t_{,T}^{2}f(1-V^{2})=\frac{r_{,T}^{2}(1-V^{2})}{fV^{2}}\text{,}  \label{F}
\end{equation}%
\begin{equation}
G=\frac{t_{,\rho }^{2}f(1-V^{2})}{V^{2}}=\frac{r_{,\rho }^{2}(1-V^{2})}{f}%
\text{.}  \label{G}
\end{equation}

Eqs. (\ref{rv}), (\ref{F}) correspond to eqs. (9), (10) of \cite{fom}.

To ensure that the left hand side of (\ref{dt}) and (\ref{dr}) is the total
differential, the integrability conditions should be satisfied:

\begin{equation}
\left( \frac{e_{0}}{f}\sqrt{F}\right) _{,\rho }=-\left( \frac{\sqrt{G}P_{0}}{%
f}\right) _{,T}  \label{int1}
\end{equation}%
\begin{equation}
\left( e_{0}\sqrt{G}\right) _{,T}=-\left( P_{0}\sqrt{F}\right) _{,\rho }
\label{int2}
\end{equation}

Here,%
\begin{equation}
f_{,\rho }=e_{0}\sqrt{G}f^{\prime }(r)\text{,}  \label{fro}
\end{equation}%
\begin{equation}
f_{,T}=-P_{0}\sqrt{F}f^{\prime }(r)  \label{fT}
\end{equation}

After substitution into (\ref{int1}) we get%
\begin{equation}
-\sqrt{FG}f^{\prime }(r)+\left( e_{0}\sqrt{F}\right) _{,\rho }+\left( P_{0}%
\sqrt{G}\right) _{,T}=0  \label{int1a}
\end{equation}

If $F=1$, (\ref{T}) coincides with (\ref{tmp}). In general, $%
e_{0}=e_{0}(\rho $, $T)$. Then, it cannot be interpreted as a conserved
energy, although one can define the quantity $V$ formally according to (\ref%
{e0}).

If we assume that $e_{0}$ is finite (at least, finite near the horizon), it
follows from (\ref{e0}) the universal behavior of $V:$%
\begin{equation}
V=\sqrt{1-\frac{f}{e_{0}^{2}}}\approx 1-\frac{f}{2e_{0}^{2}}\approx 1-\frac{%
\kappa }{e_{0}^{2}}(r-r_{+})\text{,}  \label{kappa}
\end{equation}%
where we took into account that%
\begin{equation}
f\approx 2\kappa (r-r_{+}),
\end{equation}%
$\kappa $ is the surface gravity that agrees with eqs. (21), (22) of \cite%
{fom}.

It is instructive to analyze the example suggested by Fomin for the
Schwarzschild metric, when $V=\tanh \frac{t}{2r\,+}$. In this case,%
\begin{equation}
e_{0}=\sqrt{1-\frac{r_{+}}{r}}\cosh \frac{t}{2r_{+}}\text{.}
\end{equation}

We see that for a fixed $r>r_{+}$, $\lim_{t\rightarrow \infty }e_{0}=\infty $%
. If, instead, we fix $t$, then $\lim_{r\rightarrow r_{+}}e_{0}=0$. For our
purposes, it is important that $e_{0}$ remain finite and nonzero for an
observer falling in a black hole. Then, we consider $t$ and $r$ as related
by equations of motion. For a geodesic observer with some $e$,%
\begin{equation}
\frac{dr}{dt}=\frac{f}{e}\sqrt{e^{2}-f}\text{,}
\end{equation}%
whence near the horizon of the Schwarzschild black hole%
\begin{equation}
\frac{t}{r_{+}}\approx \ln (\frac{r-r_{+}}{r_{+}})\text{.}
\end{equation}%
As a result, 
\begin{equation}
e_{0}\approx 1\text{.}
\end{equation}

Then, the transformations (\ref{dt}), (\ref{dr}) acquire the meaning of the
local Lorentz transformations and are equivalent to eqs. (7) of \cite{fom}.
It follows from (\ref{dt}), (\ref{dr}) that%
\begin{equation}
\frac{t^{\prime }}{r^{\prime }}=-\frac{P_{0}}{fe_{0}}\text{,}  \label{tr}
\end{equation}%
where prime denotes derivative with respect to $\rho $. Eq. (\ref{tr})
corresponds to eq. (17) of \cite{bron}. It can be also rewritten in the form%
\begin{equation}
\frac{t^{\prime }}{r^{\prime }}=-\frac{V}{f}\text{.}
\end{equation}

\section{Synchronous system and relation to BDG}

Now, we will consider a particular case when $e_{0}$ does not depend on $T$.
Then, it follows from (\ref{int1a}) with (\ref{int1}) with (\ref{fro}), (\ref%
{fT}) that%
\begin{equation}
\frac{de_{0}}{d\rho }\sqrt{F}+e_{0}\left( \sqrt{F}\right) _{,\rho }-\frac{1}{%
2}\sqrt{FG}f^{\prime }(r)+P_{0}\left( \sqrt{G}\right) _{,T}=0.  \label{1}
\end{equation}

And, (\ref{int2}) gives us%
\begin{equation}
e_{0}\left( \sqrt{G}\right) _{,T}+P_{0}\left( \sqrt{F}\right) _{,\rho }+%
\frac{1}{P_{0}}\sqrt{F}e_{0}\frac{de_{0}}{d\rho }-\frac{1}{2}\frac{e_{0}}{%
P_{0}}\sqrt{FG}f^{\prime }(r)=0.  \label{2}
\end{equation}

From these two equations we obtain that $F_{,\rho }=0$, so eqs. (\ref{1})
and (\ref{2}) are equivalent. If $F=F(T)$, we can always rescale time to
achieve $F=1$. Then, the frame becomes synchronous. The function $G$ should
obey the equation%
\begin{equation}
P_{0}\left( \sqrt{G}\right) _{,T}=(\frac{f^{\prime }\sqrt{G}}{2}-\frac{de_{0}%
}{d\rho })\text{,}  \label{de}
\end{equation}%
whence%
\begin{equation}
\sqrt{G}=\frac{P_{0}}{e_{0}}\mu (\rho ,r).  \label{gm}
\end{equation}

It follows from (\ref{p0}) and (\ref{fT}) that%
\begin{equation}
\left( \sqrt{G}\right) _{,T}=\frac{f^{\prime }}{2e_{0}}\text{.}  \label{gt}
\end{equation}%
After substitution in (\ref{de}) we obtain%
\begin{equation}
P_{0}^{2}\mu _{,T}=-e_{0}^{\prime }e_{0}\text{.}
\end{equation}%
Then, it is easy to find the solution with the help of the ansatz%
\begin{equation}
\mu =z(\rho )+e_{0}e_{0}^{\prime }\eta (r)\text{,}
\end{equation}%
where $\eta ^{\prime }=P_{0}^{-2}$, whence%
\begin{equation}
\eta =\int^{r}\frac{d\bar{r}}{P_{0}^{3}(\bar{r})}
\end{equation}%
for a given $\rho $. It follows from (\ref{gm}) that%
\begin{equation}
\sqrt{G}=\frac{P_{0}}{e_{0}}[z(\rho )+e_{0}e_{0}^{\prime }\eta (r)],
\label{gmn}
\end{equation}%
so%
\begin{equation}
ds^{2}=-dT^{2}+\left( \frac{P_{0}}{e_{0}}\right) ^{2}[z(\rho
)+e_{0}e_{0}^{\prime }\eta (r)]^{2}d\rho ^{2}+r^{2}d\omega ^{2}\text{.}
\label{s1}
\end{equation}

It can be written also in the form%
\begin{equation}
ds^{2}=-dT^{2}+\left( \frac{r_{,\rho }}{e_{0}}\right) ^{2}d\rho
^{2}+r^{2}d\omega ^{2}  \label{lem}
\end{equation}%
where we used (\ref{dr}).

This exactly corresponds (in our notations) to eqs. (19), (20) of \cite{bron}%
.

Thus we obtained the synchronous form of the metric from the local Lorentz
transformation following the approach of \cite{fom}. Meanwhile, it was found
in \cite{bron} due to analysis of equations of motion of geodesics with
different energies.

If the requirement $\left( \frac{\partial e_{0}}{\partial T}\right) _{\rho
}=0$ is relaxed, the metric function depends, in general, on both $T$ and $%
\rho $. Then, world lines of fiducial observers with $\rho =const$ and $%
\theta =const$, $\phi =const$ are not geodesics. Indeed, in this case we
have for the nonzero component of the four-acceleration $a^{\mu }$:%
\begin{equation}
a^{\rho }=\frac{F_{,\rho }}{2FG}\text{,}
\end{equation}%
\begin{equation}
a^{2}=a_{\mu }a^{\mu }=\frac{\left( F_{,\rho }\right) ^{2}}{4GF^{2}}.
\end{equation}

As it is assumed, by construction, that $F$ and $G$ are finite and nonzero
on the horizon, acceleration $a$ remains finite there.

If, instead of $T$ and $\rho $, one uses $T$ and $r$, the generalization of
the GP frame can be obtained. Indeed, it follows from (\ref{dt}), (\ref{dr})
that%
\begin{equation}
ds^{2}=-\frac{fF}{e_{0}^{2}}dT^{2}-2\frac{P_{0}}{e_{0}^{2}}FdTdr+\frac{%
Fdr^{2}}{e_{0}^{2}}.  \label{fg}
\end{equation}

Obviously, metric (\ref{fg}) is regular in the vicinity of the horizon. If $%
e_{0}=const$, $F=1$, we return to (\ref{gp}) after change $T\rightarrow -T$.

\section{Double GP coordinates}

In Ref. \cite{jose}, two coordinates $\tilde{t}$ and $\tau $ were used for
constructing a regular Schwarzschild metric. These coordinates represent
advanced and retarded GP coordinates. In this Section, we extend the
corresponding procedure considering more general metrics (\ref{met}). It is
quite straightforward but, bearing in mind that corresponding formulas can
be useful in further applications, we list them explicitly. Let us introduce
the coordinate $\tau $ according to%
\begin{equation}
d\tau =e_{0}dt-e_{0}\frac{dr}{f}V\text{.}
\end{equation}

Then, 
\begin{equation}
ds^{2}=\frac{f}{4P_{0}^{2}e_{0}^{2}}[f(d\tilde{t}^{2}+d\tau
^{2})]-2(2e_{0}^{2}-f)d\tau d\tilde{t}]+r^{2}d\omega ^{2}.  \label{tt}
\end{equation}

This metric is still deficient near the horizon. To repair this, one can
introduce Kruskal-type variables $\tilde{t}^{\prime }$ and $\tau ^{\prime }$%
. Let, for simplicity, $e_{0}$ be constant. Then,%
\begin{equation}
\tilde{t}^{\prime }=t_{0}\exp (\frac{\kappa \tilde{t}}{e_{0}})=t_{0}\exp
(\kappa t+\kappa \frac{\chi }{e_{0}})\text{,}  \label{t1}
\end{equation}%
\begin{equation}
\tau ^{\prime }=-t_{0}\exp (-\frac{\kappa \tilde{t}}{e_{0}})=-t_{0}\exp
(-\kappa t+\kappa \frac{\chi }{e_{0}})\text{,}  \label{tau1}
\end{equation}%
where $t_{0}$ is some constant,%
\begin{equation}
\chi =\int \frac{dr\sqrt{e_{0}^{2}-f}}{f}=\int dr^{\ast }\sqrt{e_{0}^{2}-f}%
=e_{0}\int dr^{\ast }V\text{,}
\end{equation}%
$\kappa $ is the surface gravity, $r^{\ast }$ is defined in (\ref{tort}).
Then, one can check that in variables $\tilde{t}^{\prime }$, $\tau ^{\prime
} $ the metric takes the form 
\begin{equation}
ds^{2}=\frac{f}{4P_{0}^{2}e_{0}^{2}\kappa ^{2}}[f(\frac{d\tilde{t}^{\prime 2}%
}{\tilde{t}^{\prime 2}}+\frac{d\tau ^{\prime 2}}{\tau ^{\prime 2}}%
)]+2(2e_{0}^{2}-f)\frac{d\tau ^{\prime }}{\tau ^{\prime }}\frac{d\tilde{t}%
^{\prime }}{\tilde{t}^{\prime }}]+r^{2}d\omega ^{2}\text{.}  \label{mettau}
\end{equation}%
Near the horizon,%
\begin{equation}
f\approx 2\kappa (r-r_{+})\text{, }\chi =\frac{e_{0}}{2\kappa }\ln
(r-r_{+})+\chi _{reg}\text{,}
\end{equation}%
where $\chi _{reg}$ is regular near $r_{+}$. Then,%
\begin{equation}
r^{\ast }\approx \frac{1}{2\kappa }\ln (r-r_{+})\text{,}
\end{equation}%
\begin{equation}
\tilde{t}^{\prime }\approx t_{0}\exp \kappa v\text{,}
\end{equation}%
\begin{equation}
\tau ^{\prime }\approx -t_{0}\exp (-\kappa u)\text{,}
\end{equation}%
where%
\begin{equation}
u=t-r^{\ast }\text{, }v=t+r^{\ast }\text{.}
\end{equation}

Then, near the horizon, taking into account (\ref{kappa}), we have%
\begin{equation}
f\approx -2\kappa \left( \frac{\tilde{t}^{\prime }\tau ^{\prime }}{t_{0}^{2}}%
\right) \text{.}
\end{equation}

As a result, the metric (\ref{mettau}) is regular near the horizon. If $%
t_{0}=M$ and $f=1-\frac{2M}{r}$, we return to the Schwarzschild case
considered in \cite{jose}. The whole space-time splits to four regions,
similarly to the Kruskal metric in the Schwarzschild case. Transformations (%
\ref{t1}), (\ref{tau1}) correspond to the quadrant I in \cite{jose} and can
be adjusted to other quadrants. We will not dwell upon on this.

If $e_{0}=1$, we return to the standard transformations that bring the
metric into the Kruskal form. Now, we can also perform a limiting transition
\thinspace $e_{0}\rightarrow \infty $ and observe that%
\begin{equation}
u=\lim_{e_{0}\rightarrow \infty }\frac{\tau }{e_{0}}=-\int \frac{dr}{f}%
=t-r^{\ast }\text{,}
\end{equation}%
\begin{equation}
v=\lim_{e_{0}\rightarrow \infty }\frac{\tilde{t}}{e_{0}}=t+r^{\ast }\text{.}
\end{equation}%
In this limit,%
\begin{equation}
ds^{2}=-fdudv+r^{2}d\omega ^{2}\text{,}
\end{equation}%
so $u$ and $v$ have the meaning of the Eddington-Filkenstein coordinates. In
doing so, $\tilde{t}^{\prime }$ and $\tau ^{\prime }$ have the meaning of
standard Kruskal coordinates.

It is worth noting \ an important scale property of coordinates $\tilde{t}%
^{\prime }$ and $\tau ^{\prime }$. One can compare two limits: (i) $%
e_{0}\rightarrow \infty $ for any fixed $r\geq r_{+}$ and (ii) the horizon
limit $r\rightarrow r_{+}$ for any fixed $e_{0}$. In both limits these
coordinates behave in the same manner. We see that the value $e_{0}$ does
not have a crucial influence on the coordinate frame, the metric remains
regular on the horizon.

\section{Some examples}

In this section we present some examples how different metrics, initially
discovered using totally different approaches can be incorporated into the
general scheme described in the present paper.

First of all, we can note that Eq. (\ref{lem}) is generalization of the Lema%
\^{\i}tre - Tolman - Bondi solution (LBT) of Einstein equations valid for
dust. To see this, is sufficient to write 
\begin{equation}
e_{0}^{2}=1+h(\rho )\text{. }
\end{equation}%
Then, it corresponds to eq. (103.6) of \cite{LL}, where we used $h$ instead
of $f$ in \cite{LL} and $\rho $ instead of $R$. Meanwhile, we would like to
stress that the metric (\ref{lem}) is more general and, in particular, its
origination can have nothing to do with dust.

From the other hand, (\ref{lem}) \ can be considered as generalization of
the Novikov frame \cite{nov63} used for the description of the Schwarzschild
metric, if we identify $e_{0}^{2}(\rho )=\frac{R^{\ast 2}}{1+R^{\ast 2}}$ in
eq. (31.12a) of \cite{mtw}.

Another interesting example appears if we put $e_{0}=1$ and%
\begin{equation}
f(r)=1-H^{2}r^{2}.
\end{equation}

It is convenient to rescale $\rho $ in such a way that $z(\rho )=1.$

It is seen from (\ref{p0}) that%
\begin{equation}
P_{0}=Hr.
\end{equation}

Then, eq. (\ref{gmn}) gives us%
\begin{equation}
\sqrt{G}=Hrz(\rho )\text{.}
\end{equation}%
It is convenient to take $z(\rho )=1$. It follows from (\ref{dr}) that%
\begin{equation}
r_{,\rho }=Hr
\end{equation}%
and it follows from (\ref{int2}) that%
\begin{equation}
r_{,T}=-r_{,\rho }\text{.}
\end{equation}%
As a result, we can write%
\begin{equation}
r=r_{0}\exp (H\rho -HT)\text{,}
\end{equation}%
where $r_{0}$ is a constant. We see that the expression for $r$ is
factorized into a product of a function of $\rho $ and a function of $T$.
This means that by appropriate redefinition of $\rho $ in the form $\chi
=\exp (H\rho )$ we can kill all the dependence $r$ upon $\rho $ and obtain a
metric with the dependence upon $T$ only. It is convenient also to choose $%
r_{0}=H^{-1}$ and make redefinition $\tilde{T}=-T$. Then,%
\begin{equation}
ds^{2}=-dT^{2}+\frac{\exp (2HT)}{H^{2}}(d\chi ^{2}+\chi ^{2}d\omega ^{2}) 
\text{,}  \label{DS}
\end{equation}%
where we omitted tilde. This is nothing else than the standard Friedmann
form of the de Sitter flat metric. It is interesting that allowing for
non-constant $e_0$ it is possible to get also positively and negatively
curved de Sitter solutions, see \cite{bron}.

As for GP form of the metric, it has the form 
\begin{equation}
ds^2=-(1-H^2r^2)dt^2 + 2Hr drdt + dr^2
\end{equation}
from which we can extract the Hubble law for the velocity of the flow $V=Hr$%
. It is known that this form is valid not only for de Sitter solution, but
for an arbitrary Friedmann cosmology \cite{Faraoni}.

Note that the fact that the resulting diagonal metric (\ref{DS}) appears to
be a homogeneous one explicitly is connected with a particular form of the
function $f$ in eq. (\ref{p0}) and particular value $e_{0}=1$ which leads to
factorizable expression for $r$. Meanwhile, in the next section we will see
that there exists another family of homogeneous metrics existing for an
arbitrary function $f$.

\section{The limit $e_{0} \to 0$}

Let us consider the limiting transition $e_{0}\rightarrow 0$. It cannot be
done in the metric (\ref{GP}) directly. In this limit, the axis $r$ and $T$
become collinear since in (\ref{dr}) the term with $d\rho $ drops out. As a
result, these coordinates fail to be suitable for constructing a regular
frame. Also, the limit under discussion cannot be taken in the form of
metric (\ref{s1}), (\ref{lem}). Formally, the proper distance between two
arbitrary points with different values of their radial coordinate $r$ grows
like $1/e_{0}$ and the metric becomes degenerate.

However, for a synchronous metric the limit $e_{0}=0$ is allowed. To make a
meaningful result, we need to rescale the spatial coordinate according to $%
\rho =e_{0}\tilde{\rho}$ and take the limit under discussion only
afterwards. Then, it follows from (\ref{dr}) with $e_{0}=0$, $F=1$ that%
\begin{equation}
T=-\int^{r}\frac{d\bar{r}}{\sqrt{g(\bar{r})}},
\end{equation}%
where $g=-f>0$. Thus this transformation is legitimate under the horizon
only. It brings the metric in the form 
\begin{equation}
ds^{2}=-dT^{2}+g(r(T))d\tilde{\rho}^{2}+r^{2}(T)d\omega ^{2}\text{.}
\label{32}
\end{equation}

Schematically, the timelike geodesics with $e_{0}=0$ are depicted on Fig. 1
where a relevant part of the Kruskal diagram is depicted. 
\begin{figure}
\begin{center}
\includegraphics[width=0.5\textwidth]{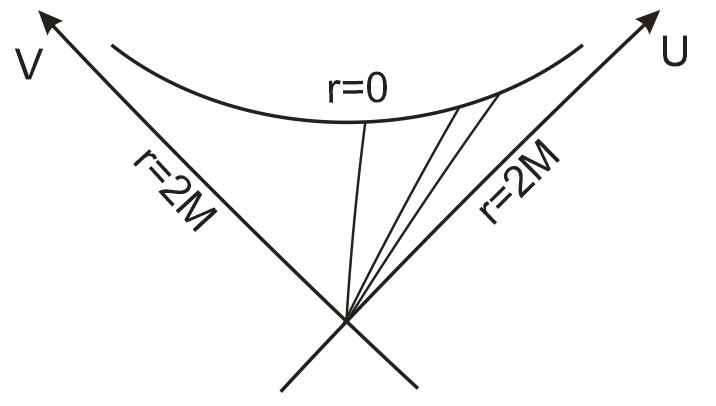}
\caption{Timelike geodesics for $e_{0}=0$.}
\label{ Kruskal}
\end{center}
\end{figure}
It is worth nothing that in synchronous form (\ref{32}) of the metric the
variable $\tilde{\rho}$ is always a spatial one, and $T$ is always a
temporal one. However, some peculiarities of the $e_{0}=0$ case lead to
peculiar properties of the corresponding synchronous frame. It can be easily
seen that the metric now depends on the temporal coordinate only, becoming
an homogeneous one. This is however not surprising, since, as the $T=const$
hypersurface coincides now with the $r=const$ hypersurface, and any spatial
dependence in a spherically symmetric metric is in fact the $r$-dependence.
Therefore, it is clear that the hypersurface $T=const$ in the $e_{0}=0$ case
has no spatial dependence at all. The central singularity is not present in
the any nonsingular $T=const$ plane, and instead, is present in the
observer's future.

This form of the metric can also be obtained directly from (\ref{met}) if
one interchanges the role of coordinates $r$ and $t$ and makes the
coefficient $g_{\tilde{t}\tilde{t}}=-1$ by rescaling the time coordinate.
This is just the form, first introduced by Novikov - see \cite{nov61} and
eqs. 2.4.8 and 2.4.9 in \cite{fn}. It can be considered as particular case
of the cosmological Kantowski-Sachs metric.

The cosmological interpretation of this metrics gives a non-formal
explanation of a curious fact about time needed to reach a singularity from
a horizon. Indeed, the coordinate time before cosmological singularity $%
\Delta T=r_{+}$ obviously does not depend on a particular motion of an
observer. As for the proper time from the horizon crossing to singularity
hitting, it differs from $\Delta T$ by a Lotentz factor originating from the
relative motion of the object in question with respect to the $e_{0}=0$
frame. As it is known from the SR, the Lorentz factor can only make the
proper time shorter, so $\Delta T$ is the maximum possible proper time from
a horizon crossing to a singularity hitting, and it is achieved if the
observer moves along the geodesic with $e_{0}=0$ -- see \cite{our} for
detail, where other formal and informal treatments of this question have
been given.

Returning to the GP metric (\ref{GP}), we can note that despite the original
GP coordinate system has no smooth $e_{0}\rightarrow 0$ limit, we can easily
write down another coordinate system with a smooth limit at $e_{0}=0$.
Indeed, if instead of $\tilde{t}$ and $r$, one uses coordinates $\tilde{t}$
and $t$, then, after substitution of (\ref{tmp}) into (\ref{GP}), we obtain
the \ metric in the form%
\begin{equation}
ds^{2}=-\frac{g}{P_{0}^{2}}d\tilde{t}^{2}+\frac{g^{2}dt^{2}}{P_{0}^{2}}+%
\frac{2ge_{0}}{P_{0}^{2}}dtd\tilde{t}+r^{2}(\tilde{t})d\omega ^{2}\text{,}
\end{equation}%
$g=-f>0$ under the horizon It can be rewritten in the form%
\begin{equation}
ds^{2}=-d\tilde{t}^{2}+\frac{g^{2}}{P_{0}^{2}}(dt+\frac{d\tilde{t}e_{0}}{g}%
)^{2}+r^{2}(\tilde{t})d\omega ^{2}.  \label{g}
\end{equation}

As under the horizon the coordinate $t$ is space-like, the metric is
expressed through one space-like and one time-like coordinate (in contrast
to the original GP which has two time-like coordinates under the horizon).
The non-diagonal term defines a coordinate "flow velocity" $-\frac{e_{0}}{g}$
which can be interpreted as a velocity with respect to $e_{0}=0$ frame.
Indeed, in the $e_{0}=0$ limit it vanishes. It is known that the 3-velocity
with respect to $e_{0}=0$ frame of a radially falling particle with the
energy $e$ is equal to $-e/P$ (see eq. (97) in \cite{we}). We get this value
from the coordinate velocity if we remember that physical distance interval $%
dl$ is connected with the interval of the space-like coordinate $dt$ through 
$dl=(g/P)dt$.

So that, this metric, in some sense dual to GP, has better behavior under
the horizon than the original GP and allows a smooth transition to $e_{0}=0$
limit.

\section{Summary}

Thus we established the connection between two kinds of approaches, both of
them being connected with the particle dynamics through the parameter $%
e_{0}\,.$ In this sense, we revealed the meaning of main coordinate
transformations from the original metric$.$ Outside the horizon, some
results are known but we extended corresponding interpretation, having
considered the region inside the horizon.

Previous papers showed how to unify separate metrics and transformations. We
made a next step and showed how one can unify the whole classes of unifying
transformation. Namely, if the parameter of coordinate transformation $%
e_{0}=e_{0}(\rho )$, the Fomin metric (\ref{fom}) turns into the BDG frame.
It is worth noting that the metric (\ref{fom}) is more general than the BDG
one in that the coordinate lines of observers with $\rho =const$ are not
necessarily geodesics.

It is also shown that, when $e_{0}\rightarrow 0$, (\ref{fom}) turns smoothly
into the metric considered by Novikov \cite{nov61}. To the best of our
knowledge, existence of this limit was not considered before in the context
of black hole metric under the horizon. Thus the coordinates frame such as
the Kruskal-Szekeres, homogeneous Kantowski-Sacks metric inside the horizon
and Lema\^{\i}tre ones, which look so differently, are now united as
elements of a whole picture.

By contrary, the generalized GP metric has no smooth \thinspace $%
e_{0}\rightarrow 0$ limit. In a sense, we proposed a metric which can be
considered as dual to GP. This new form of metric has a good behavior under
the horizon, in particular, it is regular for $e_{0}=0$.

It is of interest to try extension of the approach under discussion to the
rotating case. Especially interesting in this context is the possibility to
buid a general approach that would unite the coordinate transformations to
regular frames with the the Janis-Newman algorithm \cite{jan} that relates
static solutions and rotating metrics. Also, it is of interest to generalize
the approach under discussion to higher dimensions. All this requires
separate treatment.

\section{Acknowledgement}

This paper has been supported by the Kazan Federal University Strategic
Academic Leadership Program. AT has been supported by the Interdisciplinary
Scientific and Educational School of Moscow University in Fundamental and
Applied Space Research.

\end{document}